\documentclass[aps,prl,superscriptaddress,reprint]{revtex4-1}
\UseRawInputEncoding
\usepackage{amsmath,amssymb}                                                                                                      
     \usepackage{graphicx}                                                                                                             
                                                                                                                                       
     \usepackage{float}                                                                                                                
     \usepackage[usenames,dvipsnames]{xcolor}                                                                                          
     \usepackage{amsthm,comment}                                                                                                       
     \usepackage{booktabs}                                                                                                             
     \usepackage[export]{adjustbox}                                                                                                    
     \usepackage[section]{placeins}                                                                                                    
     \usepackage[caption=false]{subfig}                                                                                                
     \usepackage[percent]{overpic}                                                                                                     
     \bibpunct{[}{]}{,}{n}{}{}                                                                                                         
     \definecolor{red1}{HTML}{FF4136}                                                                                                  
     \definecolor{green1}{HTML}{00802b}

    \usepackage[hidelinks]{hyperref} 

\hypersetup{colorlinks=true,citecolor=Red,linkcolor=Blue,urlcolor=Black}
    \usepackage[export]{adjustbox}                                                                                                    

    \graphicspath{{Figures/}}

\begin{document}
\title{Non-trivial fusion of Majorana zero modes in interacting quantum-dot arrays}

\author{Bradraj Pandey}
\affiliation{Department of Physics and Astronomy, The University of 
Tennessee, Knoxville, Tennessee 37996, USA}
\affiliation{Materials Science and Technology Division, Oak Ridge National 
Laboratory, Oak Ridge, Tennessee 37831, USA}
\author{Satoshi Okamoto}
\affiliation{Materials Science and Technology Division, Oak Ridge National 
Laboratory, Oak Ridge, Tennessee 37831, USA}
\author{Elbio Dagotto}
\affiliation{Department of Physics and Astronomy, The University of 
Tennessee, Knoxville, Tennessee 37996, USA}
\affiliation{Materials Science and Technology Division, Oak Ridge National 
Laboratory, Oak Ridge, Tennessee 37831, USA}

\date{\today}

\begin{abstract}

Motivated by recent experimental reports of Majorana zero modes (MZMs) in quantum-dot systems 
at the ``sweet spot'', where the electronic hopping $t_h$ is equal to the superconducting coupling $\Delta$, 
we study the  time-dependent spectroscopy corresponding to the {\it non-trivial} fusion of MZMs. 
The expression non-trivial refers to the fusion of Majoranas from different original pairs of MZMs, each with well-defined parities. 
For the first time, we employ an experimentally accessible time-dependent real-space local density-of-states (LDOS) method
 to investigate the non-trivial MZMs fusion outcomes in canonical chains and in a  Y-shape array of interacting electrons.
In the case of quantum-dot chains where two pairs of MZMs are initially disconnected, after fusion 
we find equal-height peaks in the electron and hole components of the LDOS, 
signaling non-trivial fusion into both the vacuum $I$ and fermion $\Psi$ channels with equal weight. 
For $\pi$-junction quantum-dot chains, where the superconducting phase has opposite signs on the left and right portions of the chain, 
after the non-trivial fusion, we observed the formation of an exotic two-site MZM  
near the center of the chain, coexisting with another single-site MZM. 
Furthermore, we also studied the fusion of three MZMs in the Y-shape geometry. 
In this case, after the fusion we observed the novel formation of another exotic multi-site MZM,
with properties depending on the connection and geometry of the central region of the Y-shape quantum-dot array.
\end{abstract}
\maketitle

\section{I. Introduction}
Majorana zero modes are attracting much attention due to their potential application in developing 
fault-tolerant quantum computation~\cite{Kitaev2,Sarma,Nayak}. The Majorana zero modes (MZMs) follow non-Abelian statistics 
and allow the non-local encoding of quantum information, which makes MZMs good candidates to utilize 
as qubits in topological quantum computations ~\cite{Kitaev1,Nayak,Shnirman}. Recently, in coupled quantum-dot systems, a pair of 
localized MZMs were observed in the tunneling conductance measurements at the sweet spot $t_h=\Delta$, where 
the electronic hopping $t_h$ and superconducting coupling $\Delta$ are equal in magnitude~\cite{Dvir}. 
These quantum-dot systems~\cite{Jay,Souto,Mills,Deng,Loss} allow to realize the idealized Kitaev chain with 
gate-tunable experimental parameters~\cite{Liu,Wang,Wang1}. Realizing MZMs via quantum dots significantly reduces 
the problem of formation and detection of the MZMs, 
as compared to the more standard proximitized semiconducting nano-wire systems which are
affected by random disorder~\cite{Jay,Pan}.

This recent experimental progress in quantum-dot systems provides a platform to test the 
non-Abelian statistics of Majorana fermionic candidates~\cite{Dvir,Mazur}. Fusion and braiding are two fundamental 
characteristics of non-Abelian anyons~\cite{Aasen,Alicea}. The realization of MZMs at the sweet spot allows the 
study of the fusion and braiding of MZMs even in small systems~\cite{Dvir} (because in this case the MZMs are fully localized at a single site), 
as compared to the semiconducting nanowires that need a more extensive system. The sweet spot also facilitates 
analytical calculations for special cases~\cite{Leijnse,pandey1}. The fusion of MZMs and detection of their outcomes 
in quantum-dot experiments are expected to be easier than performing braiding of MZMs in other platforms.

The multiple fusion outcomes of MZMs, $\gamma \times \gamma = I + \Psi$,  are related to their non-Abelian
statistics~\cite{Nayak},
because two MZMs ($\gamma$) after fusion can result in either vacuum $(I)$ or a fermion $(\Psi)$~\cite{Aasen}.
However, the fusion of MZMs can be designed in two ways, namely a ``trivial$"$ and a ``non-trivial$"$ procedure. 
In the trivial case, the fusion outcome is deterministic (either 100$\%$ $I$ or  100$\%$ $\Psi$),
as the fusion of MZMs occurs within the same pair with well-defined parity +1 or -1. This trivial fusion can be performed using
just one pair of MZMs in a chain, by moving one edge MZM towards the other edge MZM~\cite{pandey2}.
On the other hand,  the non-trivial fusion refers to the fusion of Majoranas belonging to different initially disconnected 
pairs of MZMs each with pre-defined parities. In this non-trivial case, 
the fusion outcomes  can yield both vacuum $I$ and regular fermion $\Psi$ with equal probability 50$\%$~\cite{Aasen,Zhou}.
In order to perform non-trivial MZMs, we need at least 
two-pairs of MZMs, i.e. at least four MZMs.  This paper mainly focuses 
on the time-dependent non-trivial fusion using two and three pairs of MZMs in
models simulating interacting quantum-dot systems.

\begin{figure}[!ht]
\hspace*{-0.5cm}
\vspace*{0cm}
\begin{overpic}[width=1.0\columnwidth]{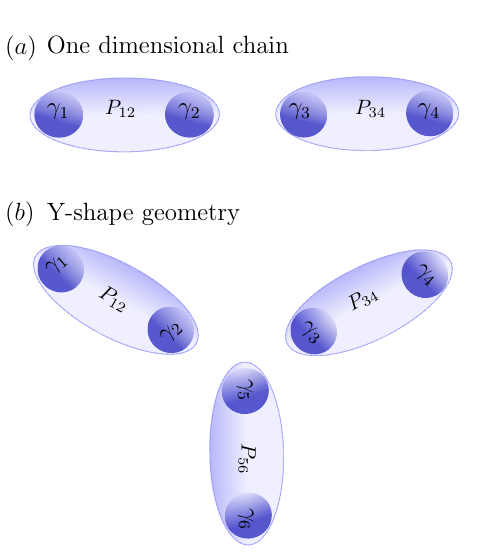}
\end{overpic}
\caption{Non-trivial fusion of Majoranas. (a)  Schematic representation of two pairs of Majorana zero modes ( [$\gamma_1$, $\gamma_2$] and [$\gamma_3$, $\gamma_4$]) in a quantum dot array. 
At time $t=0$, the parities of the left $P_{12}=-i\langle \gamma_1 \gamma_2 \rangle$ and right $P_{34}=-i\langle \gamma_3 \gamma_4 \rangle$ pairs  of MZMs are well defined. 
There is no hopping and no pairing coupling between the Majoranas  $\gamma_2$ and $\gamma_3$.  
(b) Schematic representation of the three-pairs of Majoranas zero modes in $Y$-shape geometry.
 At time $t=0$, the parities of three pairs of MZMs $P_{12}=-i\langle \gamma_1 \gamma_2 \rangle$, $P_{34}=-i\langle \gamma_3 \gamma_4 \rangle$ 
and  $P_{56}=-i\langle \gamma_5 \gamma_6 \rangle$  are well defined. Initially, there is no hopping and  pairing coupling between the three central Majoranas $\gamma_2$,$\gamma_3$ and $\gamma_5$. 
For non-trivial fusion, the time-dependent hopping and pairing amplitude between the different pairs of MZMs   
were varied with time.}
\label{fig1}
\end{figure}

 For the more canonical semiconductor nanowire setups, the detection of the fusion 
outcome of MZMs has been proposed by charge sensing
 based on dynamical Bogoliubov-de Gennes simulations~\cite{Aasen,Han,Bai}.
Recently, in the context of coupled quantum dots, the detection of the fusion of MZMs has been suggested 
using the parity readout of the systems~\cite{Liu1}. 
Even without fusing the MZMs, by preparing the two pairs of MZMs 
in two different ways, the testing of fusion outcome has been proposed 
by observing the fermionic parity readout (deterministic or probabilistic)~\cite{Liu1}.
However, some studies indicate that ``probabilistic parity measurements'' can also occur for trivial low-energy modes~\cite{Clarke,Danon}. 
Consequently, in the presence of low-energy modes, probabilistic parity measurements may yield false-positive signals in fusion experiments. 
Without careful manipulation of system parameters, these fusion experiments are not conclusive proof of the non-Abelian
statistics of MZMs~\cite{Clarke}.
Compared to previous theoretical studies, here for the first time we propose detecting fusion outcomes using the 
time-dependent real-space  electron and hole components of the local density-of-states methods~\cite{pandey2} 
in both canonical chains and in Y-shape arrays of interacting quantum dots.
The quantum-dot system provides a very well-controlled setup, where the local control of individual quantum dots significantly 
reduces the damaging effect of disorder in detecting the MZMs~\cite{Mazur}.
The total  local density-of-states (LDOS($\omega,t$)) should be experimentally accessible via tunneling conductance measurements 
in the existing quantum-dot setups ~\cite{Dvir,Wimmer}.

Motivated by the recent experimental realization of a minimal Kitaev chain in quantum dots 
coupled by a short superconductor-semiconductor hybrid (SC-SM)~\cite{Dvir,Bordin},
here we study the non-trivial fusion of MZMs in quantum-dot arrays at the sweet spot ($t_h=\Delta$).
 In the quantum dot experiments, the hopping and superconducting coupling between the quantum dots are
tunable by changing the electrostatic gate~\cite{Liu,Wang,Moor,Michael}.
In this work, to observe the time-dependent non-trivial fusion, 
we tune the time-dependent hopping and superconducting coupling between quantum-dot arrays, where two different pairs of MZMs exist with pre-defined parities.
We implement the time-dependent exact-diagonalization method using all the many-body states of interacting electrons 
of finite-size systems to study the spectroscopy of the non-trivial fusion of MZMs~\cite{pandey2}.
In the case of two one-dimensional chains with two pairs of MZMs (see Fig.~\ref{fig1}{\bf a}), we find equal height peaks in the electron 
and hole components of the LDOS($\omega,t$), showing the formation of both electron $\Psi$ and vacuum channels.
Surprisingly, due to the non-equilibrium effects and parity conservation of the time-evolving many-body state,
we find the equal magnitude of electron and hole peaks at both $\pm \omega$  energies in LDOS($\omega,t$).
In contrast to previous studies, for the first time we  discuss the effect of repulsive Coulomb interaction on non-trivial fusion.
 We also present the time-dependent non-trivial fusion using                                                     
  a pair of quasi-MZMs away from the sweet spot and for a pair of MZMs at the sweet spots.
  In this scenario, namely when the left portion is away from the sweet spot and the right portion is at the sweet spot, 
we observe unequal peak heights in the electron and hole components of LDOS($\omega,t$) 
  at $\pm \omega$ as time increases.

For the $\pi$-junction, where the two pairs of MZMs are initialized  with opposite
signs of the pairing amplitude after the fusion, we find that one of the central MZMs 
remains unaffected, while the second single-site  MZM  is transformed to a two-site
non-local MZM. In fact, we find that this two-site MZM is formed after {\it tunneling} 
through the centrally localized one-site MZM. The tunneling of {\it half} of the second MZM 
through another centrally localized one-site MZM is a novel effect in a strictly one-dimensional
geometry~\cite{Ching}.

Furthermore, we study the time-dependent fusion of an odd number (three) of Majoranas, 
where the three pairs of MZMs are initialized in a $Y$-shape geometry (see Fig.~\ref{fig1}{\bf b}) .
Surprisingly, during the fusion process we find zero energy peaks in the  LDOS($\omega,t$) for three different central sites
 in addition to the electron and hole peaks at finite energies. 
After the fusion of Majoranas, we find the formation of an exotic multi-site MZM. 
Interestingly, the nature of multi-site MZMs depends on the connection and direction of the couplings near the center,
which joins the legs of a $Y$-shape quantum-dot array. 

Last but not least, our results provide supportive evidence of the non-Abelian statistics of Majorana zero modes
by probing the LDOS($\omega,t$) of non-trivial fusion. 
The most conclusive and definitive method to verify their non-Abelian statistics would be through $``$braiding'' procedures~\cite{Clarke}.
Here, we are simply adapting to the novel experimental setups using quantum dots where peaks in the {\it conductance}
are associated with Majoranas. These conductance peaks would appear in our setup as peaks at zero frequency in the local density-of-states.
Future work will fully clarify the braiding procedure in the quantum dot setups, both experimentally and theoretically.

The organization of the manuscript is as follows. Section $II$ contains the non-trivial fusion of
Majorana zero modes for two one-dimensional chains with two pairs of MZMs. We divide this section 
in two subsections where the two pairs of MZMs are initialized with:
(A) the same signs for the pairing amplitudes, (B) the opposite signs for the pairing amplitudes.
 Section $III$ describes the non-trivial fusion of three MZMs in a $Y$-shape geometry.
 Finally, in section $IV$ we conclude our results.

\section {II. Non-trivial fusion of MZMs in one-dimensional chain}

In this section, we will consider two  one-dimensional quantum-dot arrays  at the sweet-spot,
 with the same sign for the superconducting phase ($\phi_1=\phi_2=0$). 
These two left and right short wires are coupled through time-dependent hopping ($t_h(t)$) and pairing ($\Delta(t)$) terms,
which can be tuned by changing the gate potential adiabatically.
In the quantum-dot experiments, the hopping and pairing terms 
(and also their relative signs) between the two quantum dots are tuned by 
 modifying the properties of Andreev bound states in a superconductor-semiconductor hybrid (SC-SM)~\cite{Liu,Moor,Michael}.
These properties are controlled by an electrostatic gate connected to the SC-SM hybrid segment.
The effective Hamiltonian for the quantum-dots arrays under the approximation of using only one level per quantum dot then
becomes:
\begin{eqnarray}
        H^L = \sum_{j=1}^{l-1}\left( -t_hc^{\ {\dagger}}_{j}c_{j+1} + e^{i\phi_1} \Delta  c_j c_{j+1} + H.c. \right),
\end{eqnarray}
\begin{eqnarray}
        H^{R} = \sum_{j=l+1}^{2l}\left( -t_hc^{\ {\dagger}}_{j}c_{j+1} + e^{i\phi_2} \Delta  c_j c_{j+1} + H.c. \right),
\end{eqnarray}
\begin{eqnarray}
        H^{C}(t) = \left( -t_h(t) c^{\ {\dagger}}_{l}c_{l+1} +  \Delta(t)  c_l c_{l+1} + H.c. \right).
\end{eqnarray}
The Coulomb interaction between quantum dots is the standard:
\begin{equation}
        H^{Int} = \sum_{j=1}^{2l}\left( V n_{j} n_{j+1} \right).
\end{equation}
     \begin{figure}[h]                                                                                                                 
     \centering                                                                                                                        
     \rotatebox{0}{\includegraphics*[width=\linewidth]{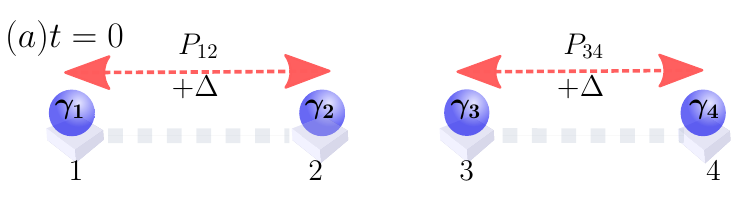}}
     \rotatebox{0}{\includegraphics*[width=\linewidth]{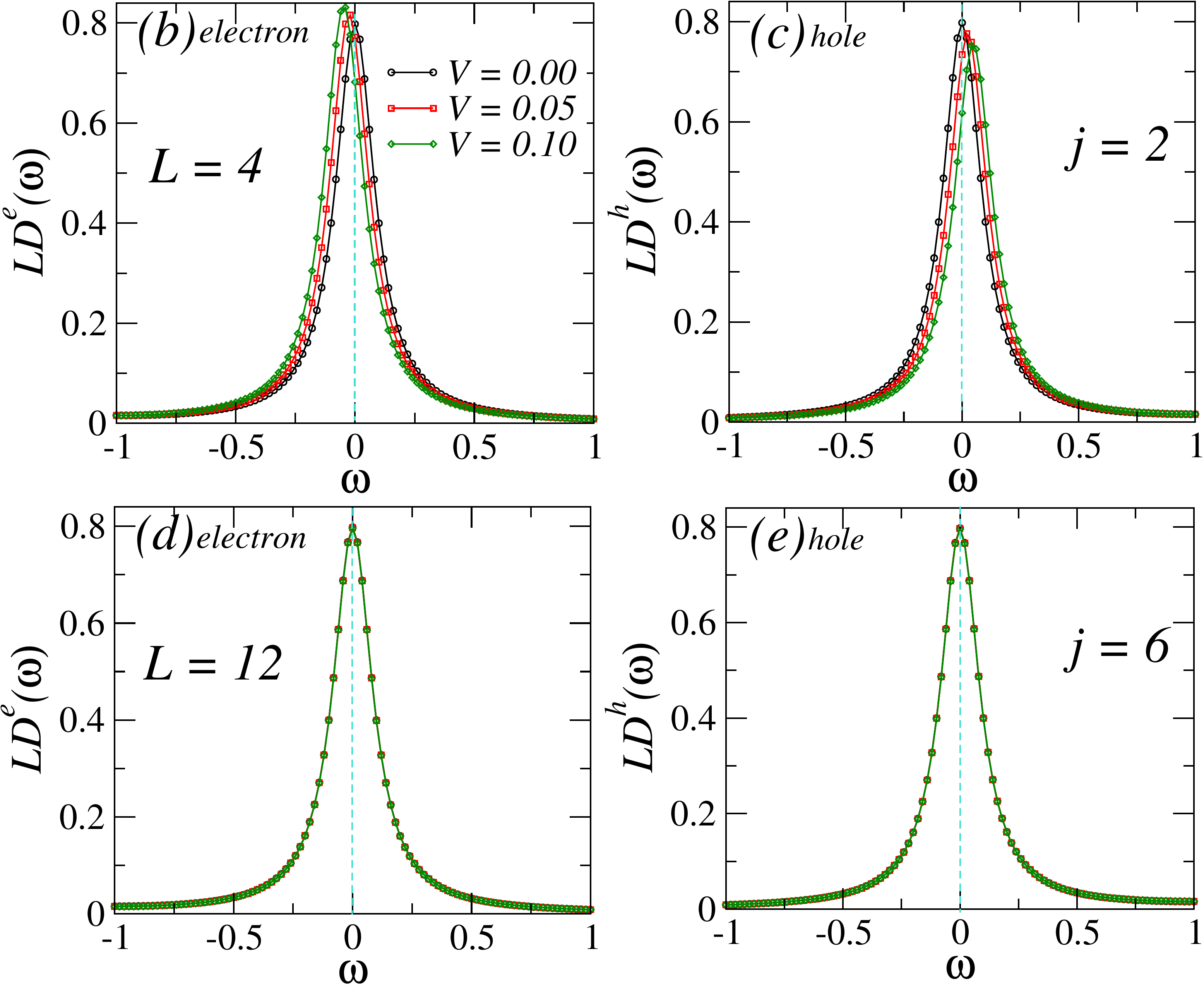}}
       \caption{(a) Schematic representation of two pairs of MZMs using a four-sites quantum-dots array with same left and right             
       superconducting phases $\phi_1=\phi_2=0$. (b,c) The electron $LD^e_j(\omega)$ and hole part $LD^e_j(\omega)$ of
       local density of states at site $j=2$ using $L=4$ sites and different values of $V$.
       (d,e) The electron $LD^e_j(\omega)$ and hole part $LD^e_j(\omega)$ of
       the local density of states at site $j=6$ using $L=12$ sites and different values of $V$.} 
     \label{fig2revised}                                                                                                                     
     \end{figure}                                                                                                                      
To perform future non-trivial fusion experiments, 
it is essential to prepare a high-quality initial state of two pairs of MZMs. 
This quality of MZMs can be compromised by the presence of disorder, interactions, or overlap 
between the MZMs within each pair involved in the fusion.
In Fig.~\ref{fig2revised}, we show the effect of repulsive Coulomb interactions $V$ on the initial state with two pairs of MZMs 
for two different system sizes $L=4$ and $12$. We calculate the electron ($LD^e(\omega)$) and hole parts ($LD^e(\omega)$) 
of the local density-of-states separately for the central sites. 
Using the eigenvectors ${|\Psi_m\rangle}$ of the Hamiltonian $H$ (Eqs.~1-4), 
the electronic component of the local density-of-states ($LD_j^e(\omega)$) for site $j$ can be written as~\cite{Herbrych,pandey2}:
   \begin{equation}                                                                                                                  
          LD_j^{e}(\omega,)= -\frac{1}{\pi} Im\left(\sum_m\frac{\left|\langle\Psi_m|c_j|\Psi_1 \rangle \right|^2}{\omega+E_m-E_1+i \eta}\right).
    \end{equation}                                                                                                                    
    For the smaller system $L=4$ (also known as Poor Man's MZMs pair), even a very small $V$ leads to a shift in the
    zero energy peaks of $LD_j^e(\omega)$ and $LD_j^e(\omega)$ from $\omega=0$ to a finite value (Fig.~\ref{fig5}(b,c)).
    This is expected as the Poor Man's MZMs lacks topological protection because of the overlap of wave functions~\cite{Danon,Mazur}.  
    Due to the smaller system size, the MZMs can hybridize if they involve more than one site, which lifts the ground state degeneracy.
    Interestingly, for the system size $L=12$ we find that the zero-energy peaks of  $LD_j^e(\omega)$ and $LD_j^h(\omega)$ 
    remains at $\omega=0$, (with equal peak heights)  for the smaller values of $V$ (see Fig.~\ref{fig2revised}(d,e)), 
    as the wave functions overlaps are negligible.
    Using the full-diagonalization method,  we find that the spectral weight at $\omega=0$ in the $LD_j^e(\omega)$ and $LD_j^h(\omega)$,  
    only arises from the four-fold degenerate ground state manifold, not from the higher-energy states above the gap. 
    These results show that long enough system sizes (but still accessible by our techniques) give topological protection against the Coulomb repulsion due to exponentially decaying wave functions, 
    and the quality of MZMs remains preserved at least for small values of $V$~\cite{Thomle,Victor}.
     In the quantum-dot experiments the Coulomb interaction $V$ can be minimized by the superconducting segment in between the quantum-dots
    ~\cite{Danon}.

\subsection{A. The case $ \phi_{\rm 1}=\phi_{\rm 2}=0$}
\begin{figure}[h]
\centering
\rotatebox{0}{\includegraphics*[width=\linewidth]{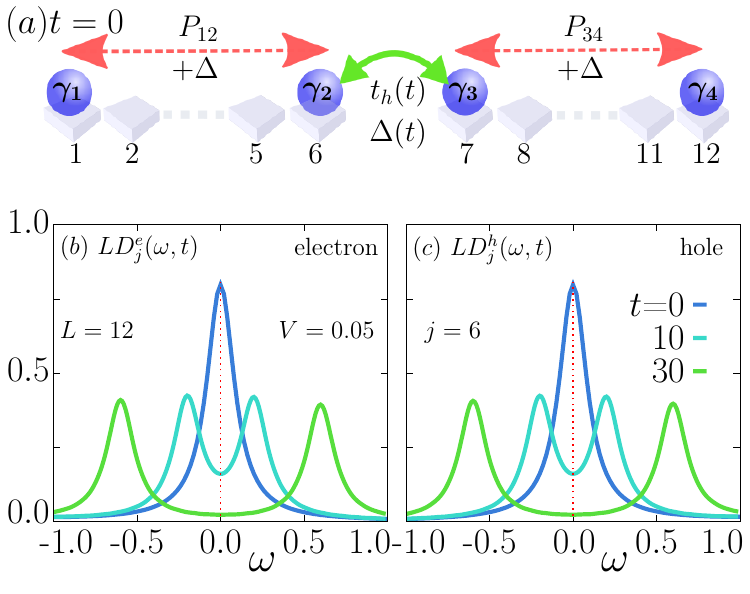}}
\caption{(a) Schematic representation of two pairs of MZMs using a 12-sites quantum dots array with same left and right 
superconducting phases $\phi_1=\phi_2=0$. At time $t=0$, 
there is no hopping and no pairing coupling between the Majoranas $\gamma_2$ and $\gamma_3$. 
The time-dependent hopping and pairing coupling between sites $j=6$ and $7$ can be established by reducing the ``barrier'' between
the left and right portions of the 12-site chain, leading to the non-trivial fusion of the central MZMs $\gamma_2$ and $\gamma_3$.
  (b) The electron component  $LD^e(\omega,t,j)$ (c) hole component $LD^h(\omega,t,j)$
at site $j=6$ and for different times $t$, at $V=0.05$. 
}     
\label{fig2}
\end{figure}

 At time $t=0$, there is no hopping and superconducting coupling between the two wires. 
As shown in Fig.~\ref{fig2}{\bf a}, the system has two separate pairs of MZMs  
$\left(\gamma_1, \gamma_2\right)$ and  $\left(\gamma_3, \gamma_4\right)$, with well defined parities $P_{12}=-i\langle \gamma_1 \gamma_2\rangle=-1$ and
$P_{34}=-i\langle \gamma_3 \gamma_4\rangle=-1$. 
The total parity (full system) of the initial many-body state can be calculated as $P_{tot}= e^{i\pi \sum_j n_j}$~\cite{Turner}.
The presence of four MZMs at $t=0$ results in a four-fold degenerate ground state, as  we can potentially create
two spinless fermions  by combining these four MZMs in pairs (thus we have degeneracy $2^2=4$). 
 Fusing the MZMs within the same pair (with parity $-1$ as example), namely $\gamma_1$ and  $\gamma_2$  or $\gamma_3$ and  $\gamma_4$,
 using a height-variable potential wall in between them,  results in the formation of a full electron (trivial fusion)~\cite{pandey2}.
This trivial fusion reveals only one fusion channel $\Psi$ with deterministic formation of a full electron. 
On the other hand, the non-trivial fusion of our focus in this paper is expected to produce both fusion channels and a more exotic intermediate dynamics.

 In the real experiments of fusion of MZMs, one needs to change the hopping and pairing 
amplitude as a function of time close to the adiabatic limit in a dynamical way. 
Namely, in the real experiments the outcome depends on the initial condition and on the speed of the process. 
Here, we perform time-dependent fusion of MZMs by changing the hopping according to the formula 
$t_h(t)=t_h(T)\frac{n\delta t}{\tau}$,
 where $1/\tau$ is the quenched rate, $\delta t= 0.001$ is the small time step we used,
and $n$ is the integer number of those steps, such that
the  $t_h(t)$ at sites $j=6$ and $7$ increases approximately linearly from $0$ to $1$ in a time $\tau =100$.
 At final time $t=T$, the time dependent hopping becomes equal to $t_h(T)=1$. 
For the time evolution, we chose the initial many-body state $|\Psi(0)\rangle$ with total parity $P_{tot}=+1$
(this parity is chosen because it corresponds to the ground state at nonzero $V$).  

To observe the fusion outcomes at intermediate time $t$, first, we time evolve the  initial wave function 
$|\Psi(0)\rangle$ up to time $t$,  using the time-dependent Hamiltonian $H(t)$ as: 
$|\Psi(t)\rangle= \mathcal{T} exp{\left (-i\int_{0}^{t}H(s) ds\right)}|\Psi(0)\rangle$,
        where $\mathcal{T}$ is the time ordering operator~\cite{Sen}.
        Next, we calculate the double-time Green function $G(t,t')$~\cite{Kennes},
        using the instantaneous Hamiltonian $H_f = H(t=t_f)$ at time $t=t_f$:
\begin{equation}
G^{elec}_j(t,t')= \langle \Psi(t)|c^{\dagger}_j e^{iH_ft'}c_j e^{-iH_ft'}|\Psi(t)\rangle.
\end{equation}
The time-dependent $LD^{e}_j$($\omega,t$) for the electronic part of the local-density of state is the 
Fourier transform of the local Green function at site $j$ with respect to $t'$.
Using the eigenvectors $|\phi_m\rangle$ of the instantaneous Hamiltonian $H(t)$ and $|\Psi(t)\rangle$ at time $t$,
the electronic component of $LD_j(\omega,t)$ at site $j$ can be written as $LD_j^{e}(\omega,t)$~\cite{pandey2} : 
\begin{eqnarray}
= \frac{-1}{\pi} Im\left( \sum_{m,n} \frac{\langle \Psi(t)| c^+_j|\phi_n\rangle \langle \phi_n| c_j |\phi_m\rangle \langle \phi_m|\Psi(t)\rangle }{e_n-e_m+\omega+i \eta} \right).
\end{eqnarray}
\noindent the broadening parameter was fixed to $\eta=0.1$ in the entire publication. 
 $e_n$ and $e_m$ denote eigenvalues of the instantaneous Hamiltonian $H(t)$.
Similarly, the hole component of  $LD_j(\omega,t)$ at site $j$ can be written as $LD_j^{h}(\omega,t)$~\cite{pandey2}:
\begin{eqnarray}
=\frac{-1}{\pi} Im\left( \sum_{m,n} \frac{\langle \Psi(t)|\phi_n\rangle \langle \phi_n| c_j |\phi_m\rangle   \langle \phi_m|c^{\dagger}_j|\Psi(t)\rangle }{e_n-e_m+\omega+i \eta} \right).
\end{eqnarray}

Interestingly, with increase in the couplings $t_h(t)$ and $\Delta(t)$, the electron $LD_j^{e}(\omega,t)$ and hole $LD_j^{h}(\omega,t)$
both shows equal-height sub-gap peaks close to  $\omega =\pm t_h(T)\frac{2t}{\tau}$ (in the rest of the paper, we use $t_h(T)=1$)  for $V=0.05$,
 reflecting the formation of equal amounts of electron and hole  at positive and negative values of $\omega$ [see  Figs.~\ref{fig2}{\bf (b,c)}].
The appearance of equal magnitude electron and hole components 
at both frequencies $\omega=\pm 2t/\tau$, is clearly a  non-equilibrium effect and it is influenced by the conservation of total parity of $|\Psi(t)\rangle$. 
Due to this dynamics, the time-evolving wavefunction $|\Psi(t)\rangle$ has an equal overlap with low-energy states $|\Psi_1\rangle$ and $|\Psi_4\rangle$ with same total parity $P=+1$ (see Appendix C). This allows for similar spectral weights close to  $\omega = -2t/\tau$ in $LD_j^{e}(\omega,t)$ 
 (transition from state $m=1$ to $n=4$)  and $LD_j^{h}(\omega,t)$ (from state $m=2$ to $n=3$), and also  similar spectral 
weights close to $\omega = +2t/\tau$ in  $LD_j^{e}(\omega,t)$ (resulting from a transition from state $m=3$ to $n=2$) 
 and  in  $LD_j^{h}(\omega,t)$  (from state $m=4$ to $n=1$). 

The equal superposition of two low-energy many-body states  in  $|\Psi(t)\rangle$  even for large $\tau=100$
  (we have also checked for the case $\tau=500$ and the results are the same [see Appendix D])
 is a unique property of non-trivial fusion. In the trivial fusion for the larger $\tau$, the time evolving wave-function  $|\Psi(t)\rangle$
 overlap only with one low-energy state of the instantaneous ground states manifold of the time evolving Hamiltonian. 
Introducing a time-dependent hopping $t_h(t)$ and pairing $\Delta(t)$, the fuse of the MZMs  $\gamma_2$ and $\gamma_3$ occurs 
non-trivially because the MZMs are from different pairs. 
The $t_h(t)$ and  $\Delta(t)$ allows for tunneling of a single electron (or a pair of electrons) from
the left to the right chain portions during the fusion process, 
which changes the  individual  parities of those left and right quantum-dot segments, 
but the total parity of the many-body state $|\Psi(t)\rangle$ remains the same~\cite{Ivar}. This results in the formation of  
both fermion $\Psi$ and vacuum $I$ channels, after fusion of  $\gamma_2$ and $\gamma_3$ non-trivially (see Appendix A) for more details).
At the final time $t=T$, the system has two MZMs ($\gamma_1$ and $\gamma_4$) at the left and right edge of the chain
(see Fig.~\ref{fig2}{\bf b}).

In summary, we have found an equal spectral weight for electron and hole components
in the time-dependent local density of states during the non-trivial fusion process of Majoranas
form different pairs. Interestingly the time-evolving state becomes an almost equal-linear superposition of 
two low-energy states, even for larger values of $\tau$.
  The appearance of equal height peaks  
of electron and hole  in the local-density of states signals the non-trivial nature of Majorana fermions.
 In the case of fusion involving two pairs of MZMs, where the two MZMs of the left pair are already slightly hybridized 
  (for $\Delta_1 =0.7$ and $\Delta_2=1.0$), this leads to an asymmetry in the peak heights close to $\omega= \pm 2t/\tau$ (see Appendix. A).
  In other words, there is a formation of unequal-height electron and hole peaks around $\omega= \pm 2t/\tau$, which is clearly different from the 
  case of an initial state with non-overlapping pairs of MZMs. 

\subsection{B. The case $\phi_{\rm 1}=\pi$, and  $\phi_{\rm 2}=0$}
\begin{figure*}[!ht]
\hspace*{-0.5cm}
\vspace*{0cm}
\begin{overpic}[width=2.0\columnwidth]{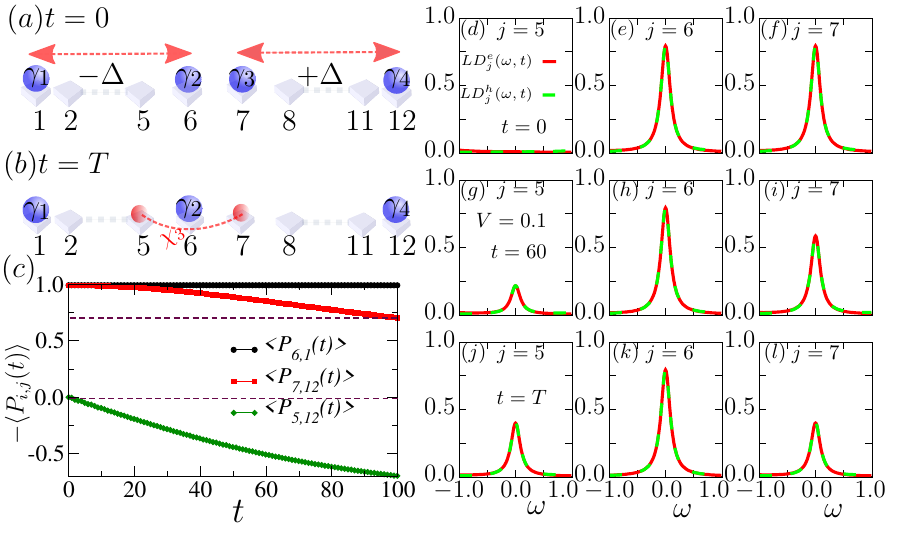}
\end{overpic}
\caption{(a) Schematic representation of two pairs of MZMs [($\gamma_1$, $\gamma_2$) and ($\gamma_3$, $\gamma_4$)]: at time $t=0$ the left part has 
	pairing coupling strength $-\Delta$, while the right part has pairing coupling strength $+\Delta$. 
	(b) Pictorial representation for the formation of a  multi-site MZM ($\chi_3$) at time $t=T$. 
	(c) Time-dependent parity $-\langle P_{i,j}(t) \rangle$ as a function of time $t$ for different pairs of Majoranas  and at $V=0.1$.
	 The electron and hole components of the time-dependent local density of state 
	$LD^e_j(t)$ and $LD_j^h(t)$ at time $t=0$  and for sites: (d) $j=5$, (e)  $j=6$, and (f) $j=7$.
	 The $LD^e_j(t)$ and $LD_j^h(t)$ at intermediate time $t=60$  and for sites: (g) $j=5$, (h)  $j=6$, and (i) $j=7$.
	 The $LD^e_j(t)$ and $LD_j^h(t)$ at the final time $t=T$  and for sites: (j) $j=5$, (k)  $j=6$, and (l) $j=7$.
}
\label{fig3}
\end{figure*}
          In this sub-section, we will study the Majorana fusion in a $\pi$-junction setup. 
In order to form a $\pi$-junction between the right and left quantum-dot arrays,
we consider two pairs of MZMs, in this case with opposite signs of 
 the pairing terms ($-\Delta$ for the left array, and $+\Delta$ for the right array).
The MZMs in the left quantum-dots array has definite parity $P_{6,1}=-i\langle \gamma_6 \gamma_1\rangle=-1$ and the 
right pair has definite parity 
$P_{7,12}=-i\langle \gamma_7 \gamma_{12}\rangle =-1$ as well (see  Fig.~\ref{fig3}{\bf a}). At $t=0$ there are no hopping and pairing terms 
between the left and right arrays (the system has a mirror symmetry with respect to a line
passing in between sites $j=6$ and $7$).

Next, we turn on the time-dependent hopping ($t_h$) and pairing $\Delta(t)$ terms (with zero phase factor as example) 
between the left and right arrays,
which effectively makes a time-dependent $\pi$-junction quantum-dot array. Note that the introduction of
time-dependent $\Delta(t)$ terms with zero phase factor breaks the initial global mirror symmetry about the 
line in between the sites $j=6$ and $7$. The choice of positive $\Delta(t)$ leads to the formation of a $\pi$-junction
at site $j=6$ (i.e. for $j\leq 6$ the pairing term is $\Delta<0$ and for $j> 6$ it has positive $\Delta >0$).
In order to observe the behavior of the central Majoranas $\gamma_2$  and  $\gamma_3$, we calculate the time-dependent 
 electron  $LD^e_j(t)$  and hole  $LD_j^h(t)$ portions of the local-density of state for sites $j=5,6$ and $7$. 
At $t=0$  the electron  $LD^e_j(t)$  and hole  $LD_j^h(t)$ portions of the 
local-density of states shows equal height peaks at $\omega=0$ for 
the sites $j=6$ and $7$, which indicate the presence of two localized MZMs [see  Figs.~\ref{fig3}{\bf(d,e,f)}].  

Increasing time to $t=60$ [ Figs.~\ref{fig3}{\bf(g,h,i)}] the peak height of $LD^e_j(t)$ and $LD_j^h(t)$
increase for site $j=5$, remain {\it constant} for site $j=6$, and  decrease  for site $j=7$.
This spectral weight shift suggests the {\it tunneling} of MZM $\gamma_3$  
from site $j=7$ to $j=5$. 
Note that the height of peaks at $\omega=0$, for the electron $LD^e_j(t)$ and hole $LD_j^h(t)$ parts of the 
local-density of states are almost equal for each site for all times $t$. 
We also find that the spectral weight at $\omega=0$, mainly arises from the low-energy sub-space of four-fold degenerate states.   
The equal-contribution of electron and hole part of the local-density of states at $\omega=0$ for these central sites,  
shows the presence of Majorana zero modes at sites $j=5,6,$ and $7$.

At the final time $t=T=100$, the  $LD^e_j(t)$ and $LD_j^h(t)$ have almost 
equal height peaks for sites $j=5$ and $j=7$ and the total spectral weights on these two sites
is close to localized on-site MZM, confirming the formation of
the multi-site MZM $\chi_3$ (see  Fig.~\ref{fig3}{\bf b}). The $LD^e_j(t)$ and $LD_j^h(t)$ for site $j=6$ remain
constant all up to time $t=T$. {\it Thus, the transfer of spectral weight from site $j=7$ to $j=5$
at $\omega=0$, hints to a tunneling effect of half of the  MZM $\gamma_3$ from site $j=7$ to 
site $j=5$, leading to formation of a multi-site MZM $\chi_3$.} This is a novel effect not reported before.

Interestingly, the low-energy four states remains degenerate, 
even after switching the time-dependent hopping $t_h(t)$ and pairing $\Delta(t)$ 
terms between the left and right arrays, showing the presence of a total of four MZMs in the system.
We also find that the time-evolving wavefunction $|\Psi(t)\rangle=u|\Psi_1\rangle+v |\Psi_4\rangle$
 becomes a superposition of two low-energy degenerate ground states with the same total parity (see Appendix C).
The amplitudes $u$ and $v$ depend on the tunneling of the  MZM $\gamma_3$ from site $j=7$ to
site $j=5$ (for $L=10$ site and time $t=20$ $|u|^2=0.8$ and $|v|^2=0.2$). 
These amplitudes $u$ and $v$ remain the same for different values of large $\tau$. 
Furthermore, to confirm the tunneling of MZM $\gamma_3$, we also calculate  the time-dependent 
parity $\langle P_{ij}(t)\rangle =- \langle \Psi(t)| \gamma_i \gamma_j|\Psi(t) \rangle $,
for different pairs of Majoranas. As shown in  Fig.~\ref{fig3}{\bf c}, the observable $\langle P_{6,1}(t)\rangle$ 
remains constant with time, showing that the MZM $\gamma_2$ remains localized at site $j=6$, 
without any change in parity of the pair ($\gamma_1, \gamma_2$). 
Meanwhile, $\langle P_{7,12}(t)\rangle $ starts decreasing and approach the value $1/\sqrt{2}$,  
at time $t=T$. On the other hand, $\langle P_{5,12}(t)\rangle $ becomes non-zero as time increases and 
approaches $-1/\sqrt{2}$.  Once again, these results suggest the formation of a  multi-site MZM with form  
$\chi_3=-\frac{1}{\sqrt{2}}\gamma_5 + \frac{1}{\sqrt{2}} \gamma_7$,
where one component of $\chi_3$ appears due to tunneling of $\gamma_3$ initially localized on site $j=7$.

For the ground-state of one-dimensional Kitaev-wire with $\pi$-junction, there are a total of four MZMs.
For the system considered in Fig.~\ref{fig3}, at the final time $t=T$,
the system forms a $\pi$-junction at site $j=6$.
To understand the behavior of the central MZMs, using just three central sites ($j=5,6,$ and $7$) near the junction,
we can write the Hamiltonian for the central region as (with $\phi_1=\pi$ and $\phi_2=0$ for the two central bonds respectively):
\begin{equation}
        H^{III} = -2i\Delta \left(\gamma^{B}_{5} \gamma^{B}_{6}+\gamma^{A}_{7} \gamma^{B}_{6} \right),
\end{equation}
where we have used the relations $c_5=\frac{1}{\sqrt{2}} e^{-i\phi_1/2}  \left(\gamma^{A}_{5}+i \gamma^{B}_{5}\right)$,
$c_{6}=\frac{1}{\sqrt{2}} \left(\gamma^{A}_{6}+i \gamma^{B}_{6}\right)$,
$c_{7}=\frac{1}{\sqrt{2}} e^{-i\phi_2/2} \left(\gamma^{A}_{7}+i \gamma^{B}_{7}\right)$. 
Interestingly, the Majorana operator $\gamma^{A}_{6}$ is absent in the Hamiltonian, 
 showing that $\gamma^{A}_{6}$ is a single-site MZM mode at site $6$. 
 The form of the Hamiltonian also suggests that one localized mode, $\gamma^{A}_{6}$,
 does not interact with any other Majorana zero mode, which could be the reason of 
 localization of the initial Majorana mode located at site $j=6$.  
Our recent work, using symmetry arguments, shows that for the two wires (each carrying a pair of MZMs) 
with a phase difference of $\pi$, the central MZMs are protected by mirror symmetry. 
The central MZMs do not fuse as they belong to different quantum numbers~\cite{pandey3}.
 After diaogonalizing the Hamiltonian $H^{III}$, 
 we find that a multi-site MZM $\chi_3=-\frac{1}{\sqrt{2}}\gamma_5 + \frac{1}{\sqrt{2}} \gamma_7$
 resides at sites $j=5$ and $j=7$ (see the SM of Ref~\cite{pandey1} for a more detail calculation). 
This ground state results indicate that out of an initial total of four single-site MZMs, there are now 
three single-site local MZMs  
(two localized at edge end sites and one at central site $j$) and one multi-site MZM with 
form $\chi_3=-\frac{1}{\sqrt{2}}\gamma_{j} + \frac{1}{\sqrt{2}} \gamma_{j+2}$.

In summary, the Majoranas near the $\pi$-junction do not fuse. 
Instead, one MZM remains a localized single-site MZM, and another transforms into a multi-site MZM 
 (located on two sites with equal amplitude). The tunneling of half of the second MZM through the
centrally localized one-site MZM  in a strict one-dimensional geometry  is an interesting dynamical effect.
 This partial tunneling of a Majorana leads to the time-evolving wavefunction in a superposition of two low-energy
degenerate states (with same total parity). 
 
\section{III. Non-trivial fusion of MZMs in $Y$-shape quantum dot arrays}
\begin{figure*}[!ht]
\hspace*{-0.5cm}
\vspace*{0cm}
\begin{overpic}[width=2.0\columnwidth]{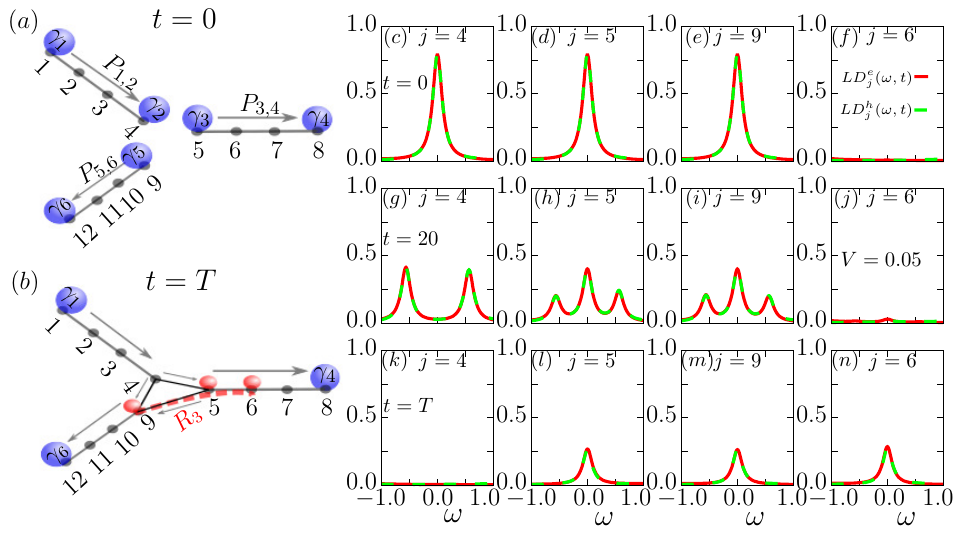}
\end{overpic}
\caption{(a) Schematic representation of three pairs of  Majorana zero modes in a $Y$-shape Kitaev wire.
At $t=0$, each pair has fixed fermionic parity $P_{1,2}=+1$,  $P_{3,4}=+1$, and  $P_{5,6}=+1$.
The direction of the arrows denote the directions of the pairing terms $\Delta$ and site index $j$  in each wire.
(b) Pictorial representation of the formation of the  multi-site MZM $R_3$ 
after the fusion of the three central MZMs at time $t=T$, after the adiabatic process ends.
          The time-dependent electron $LD^e_j(\omega,t)$ and hole  
	  $ LD^h_j(\omega,t)$ portions of the local-density of states at time $t=0$ and for sites (c) $j=4$, (d) $j=5$, (e) $j=9$, and (f) $j=6$.
          The  $LD^e_j(\omega,t)$ and $ LD^h_j(\omega,t)$ for the intermediate time $t=20$ and for sites  (g) $j=4$, (h) $j=5$, (i) $j=9$, and (j) $j=6$.
          The  $LD^e_j(\omega,t)$ and $ LD^h_j(\omega,t)$ for the final time $t=T$ and for sites  (k) $j=4$, (l) $j=5$, (m) $j=9$, and (n) $j=6$.
          These numerical calculations were performed using $L=12$ sites and $t_h=\Delta=1$, $V=0.05$.
}
\label{fig4}
\end{figure*}


This section will study the fusion of three MZMs from different pairs, further increasing the complexity of the problem. The presence of multiple Majoranas can occur
in topological materials or in quantum circuits experiments.
 The overlap between odd and even numbers of Majoranas can give different behavior in the tunneling spectra~\cite{Flensberg}. 
Here, we simulate the overlap between an odd number (three) of MZMs as a function of time , starting with fully separated MZMs.   
We consider a $Y$-shape geometry consisting of three quantum-dot chains at the sweet spot 
and with the same superconducting phase $\phi_1=\phi_2=\phi_3=0$ at each arm. 
At time $t=0$, there is no hopping and superconducting coupling between  these three quantum-dot arrays. 
As shown in  Fig.~\ref{fig4}{\bf a}, the system has three pairs of Majorana zero modes  $\left(\gamma_1, \gamma_2\right)$, $\left(\gamma_3, \gamma_4\right)$,  and 
 $\left(\gamma_5, \gamma_6\right)$, with well-defined initial parities $P_{12}=-i\langle \gamma_1 \gamma_2\rangle=+1$, 
$P_{34}=-i\langle \gamma_3 \gamma_4\rangle=+1$, and $P_{56}=-i\langle \gamma_5 \gamma_6\rangle=+1$. 
The six Majorana modes at the sweet-spot ($t_h=\Delta=1$ and $V=0$) give rise to an eight-fold degenerate ground state, 
because we can form three non-local spinless fermions
(which give rise to $2^3=8$ fold degeneracy). 
Within an eight-fold degenerate ground state, four states have individual total parity $P=+1$ and the remaining four
have parity $P=-1$. Using the four-degenerate ground state with fixed same parity one can encode two topological qubits, 
which can display all the basic operations for topological quantum computation~\cite{Pachos}. 

The Hamiltonian for the $Y$-shaped quantum-dot array (with $\phi_1= \phi_2=\phi_3=0$ at each arm) 
can be divided into four different parts. The Hamiltonian for each arm can be written as:
\begin{eqnarray}
        H^I = \sum_{j=1}^{l-1}\left( -t_hc^{\ {\dagger}}_{j}c_{j+1} +  \Delta  c_j c_{j+1} + H.c. \right),\\
        H^{II} = \sum_{j=l+1}^{2l-1}\left( -t_hc^{\ {\dagger}}_{j}c_{j+1} +  \Delta  c_j c_{j+1} + H.c. \right),\\
        H^{III} = \sum_{j=2l+1}^{3l-1}\left( -t_hc^{\ {\dagger}}_{j}c_{j+1} +  \Delta  c_j c_{j+1} + H.c. \right).
\end{eqnarray}

Moreover, in order to fuse the  MZMs from different pairs, first we switch on the time-dependent 
pairing $\Delta(t)$ and hopping $t_h(t)$ terms between each arm. 
In practice, we tune the time-dependent pairing and hopping terms as $\Delta(t)=t_h(t)=t_h(T)\frac{n\delta t}{\tau}$ 
between the central three MZMs ($\gamma_2, \gamma_3, \gamma_5$). We used $\tau=100$ and other parameters as described in the previous cases.
$T$ is the final time such that $\Delta(t)=t_h(t)=1$ at $t=T=100$. 
The time-dependent Hamiltonian coupling the arm edges is  written as:
\begin{eqnarray}
        H^{C}(t) =  -t_h (t)c^{\ {\dagger}}_{l}c^{\phantom \dagger}_{l+1} + \Delta(t)  c_l c_{l+1} + H.c. 
 \nonumber\\
                  -t_h(t)c^{\ {\dagger}}_{l+1}c^{\phantom \dagger}_{2l+1} +  \Delta(t)  c_{l+1} c_{2l+1} + H.c. 
 \nonumber\\
              -t_h(t) c^{\ {\dagger}}_{l}c^{\phantom \dagger}_{2l+1} +  \Delta(t)  c_{l} c_{2l+1} + H.c. .
\end{eqnarray}

Using the above Hamiltonian $H=H^I+ H^{II}+ H^{III}+  H^{C}(t)$, we performed the time evolution starting from the ground state with total parity $P=+1$, 
and calculated the electron  $LD^e_j(\omega,t)$ and hole  $ LD^h_j(\omega,t)$ portions of the local-density of states for various times $t$. 
At $t=0$, we find that the central edge sites $j=4, 5$, and $9$ in  $LD^e_j(\omega,t)$ and  $ LD^h_j(\omega,t)$ 
have sharp peaks at $\omega=0$ Figs.~\ref{fig4}{\bf(c,d,e,f)}.  
Introducing a time-dependent hopping and pairing term leads to the simultaneous fusion of 
three central MZMs ($\gamma_2, \gamma_3, \gamma_5$), the three belonging to different original MZM pairs. 
The fusion of central MZMs leads to a split in the initial eight-fold degenerate ground state into 
 two sets of low-energy four-fold degenerate states.  
    
As shown in  Fig.~\ref{fig4}{\bf g}, the time-dependent electron $LD^e_j(\omega,t)$ and hole  
$ LD^h_j(\omega,t)$ portions of the local-density of states at time $t=20$,
 for site $j=4$, shows peaks close to $\omega = \pm\frac{3t}{\tau}$, indicating the 
formation of both electron and hole for positive and negative frequencies. 
 Using Eqs.~6 and 7, we find the spectral weights  at $\omega= \pm\frac{3t}{\tau}$ in the electron  and hole  parts 
of $LD_j(\omega,t)$ arise from the transition  between the splitted two-set of four-fold degenerate 
states [mainly ($n=2$ to $m=6$) or  ($n=1$ to $m=5$)]. 
In fact, the time-evolving state $|\Psi(t)\rangle$ becomes a superposition of four states (two from the lower four-fold
degenerate part and rest two from the other four-fold states of the eight-fold low-energy states) of the instantaneous Hamiltonian.
 
At sites $j=5$ and $9$ [ Figs.~\ref{fig4}{\bf (h,i)}], the  $LD^e_j(\omega,t)$ and $ LD^h_j(\omega,t)$, 
shows peaks at $\omega=0$ and also close to $\omega = \pm\frac{3t}{\tau}$,
with equal spectral weight of the electron and hole portions of the time-dependent local-density of states.
These results for the sites $j=5$ and $9$ show that the  MZMs still survive for these sites and in addition there is a formation of
electron and hole close to $\omega= \pm\frac{3t}{\tau}$ due to the partial fusion of MZMs. 
Interestingly, for site $j=6$, a small peak appears at $\omega=0$ [Fig.~\ref{fig4}{\bf j}].
With further increase in time, we find for sites $j=5$ and $9$  the  peaks at $\omega=0$
  decreases with time. On the other hand for site $j=6$  the peak strength  at $\omega=0$ increases with increasing time $t$.
These results show the {\it transfer} of spectral weight from the central sites ($j=5$ and $9$) to the site $j=6$. 
For the final time after the adiabatic process, $t=T$, the $LD^e_j(\omega,t)$ and $LD^h_j(\omega,t)$ show almost equal-height peaks at $\omega=0$ 
for the three central sites $j=5,9$ and $6$, see  Figs.~\ref{fig4}{\bf (l,m,n)}. These  results indicate the formation of equal amount of
   electron and hole, and, surprisingly,  one multi-site MZM after the fusion of three central MZMs. 
In other words, we unveiled an interesting result that the fusion of three one-site MZMs leads to a single MZM spread on three sites. 

In summary, in comparison to the fusion of two MZMs (even numbers), the fusion of three MZMs (odd numbers) 
give rise to a novel multi-site MZM. The local density of states shows peaks at $\omega=0$ for three different central sites. 
Note that the multi-site MZM position depend on the pairing terms' direction. The appearance of a multi-site MZM near the central part 
is mainly due to the different coupling directions (effectively a $\pi$-junction) at the tri-junction (see Fig.~\ref{fig4}{\bf b}).

\section {IV. Summary and outlook}

In this publication, we studied the non-trivial fusion of Majorana zero modes in canonical chains, 
as well as in a Y-shape array of interacting quantum dots
close to the sweet spot in parameter space. 
We examined the real-time dynamics of the local density-of-states to
reveal the nature of the non-trivial fusion of MZMs.
The Majoranas were initialized in pairs with definite parity in separate quantum-dot arrays. 
Varying the time-dependent hopping and pairing terms between the different quantum-dot arrays, we carry out the fusion of 
Majoranas from different pairs (non-trivial fusion). We observed the fusion outcomes by calculating 
the time-dependent electron and hole part of the local-density of states. Several interesting results were unveiled:

(1) In the case of a one-dimensional chain with the same phase on each left and right wire, 
we  demonstrated the formation of both electron and hole close to $\omega=\pm 2t/\tau$ in equal magnitude for small values of Coulomb interactions.
The formation of equal height peaks of electron and hole at each $\omega=\pm 2t/\tau$ value is a dynamical effect and 
reveals the non-trivial nature of the MZMs fusion.  In fact, we find that the time-evolving states becomes an equal superposition 
of two states (with the same parities). This non-equilibrium effect is  unique in case of the non-trivial fusion.
For the trivial fusion the time-evolving states overlaps with only one state of the instantaneous Hamiltonian.
We also explored the non-trivial fusion using a pair of overlapping quasi-MZMs away from the sweet 
spot  and a pair of non-overlapping MZMs at the sweet spot. In this case, we find an asymmetry in the peak height of the
electron and hole portions of the local 
density-of-states at $\omega=\pm 2t/\tau$. The unequal-peak heights varying time in the electron and hole part 
of the local-density
arises from having an initial state with unequal contributions of the electron and hole components of the quasi-MZMs.

(2) On the other hand, quite interesting results were found for 
the case of a $\pi$-junction (with opposite phase on each left and right wire) because
the Majoranas do not fuse with one another. Instead they formed  a multi-site MZM residing on two sites near 
to an independent localized one-site central MZM.  
The time-average parity and time-dependent local-density of states reveals that
 the one-site MZM at the edge (near the center) of the left array does not fuse with other MZMs and remain localized on the same edge site. 
Surprisingly, half  the MZM of the right quantum-dot array (near the center) {\it tunnels} 
through the localized one-site MZM and forms a multi-site MZM. The tunneling of half of the MZM  even in a strict
one-dimensional geometry is a quite novel effect.
The tunneling of MZM also makes the time-evolving state become a superposition of two states from the four-fold degenerate
ground state manifold of the instantaneous Hamiltonian even for smaller quench rate. 
The amplitude of the two states in the time-evolving state depends on the amount of tunnel MZM through the central localized MZM.
In the quantum-dots system, the $\pi$-junction 
can be experimentally achieved by applying a magnetic 
flux through a superconducting loop that connects the two hybrid segments of such system~\cite{Luna}.
Additionally, the relative sign of hopping and superconducting coupling between the quantum dots can
also be altered using an electrostatic gate attached into the hybrid region, even without any application of magnetic flux~\cite{Michael}.

(3) For the fusion of MZMs in a $Y$-shape quantum-dot array, 
where the MZMs are coupled through time-dependent hopping and pairing terms
 in a triangular geometry, we show the formation of an exotic multi-site MZM after the fusion of three central MZMs from different pairs.
 Interestingly, the time-evolving state becomes a superposition of four states of instantaneous Hamiltonian
 (two from the lower four-fold degenerate and other two form higher four-fold degenerate states), due to the non-trivial fusion
and formation of multi-site MZM. In general terms, the nature and behavior of the  central multi-sites MZMs are dependent on the geometry and direction 
of the pairing terms of quantum-dot arrays.  The knowledge of the characteristics of the 
central multi-sites MZMs is important for the braiding of MZMs in dynamical and realistic settings, 
where the exchange of MZMs is performed by moving the MZMs adiabatically.
In comparison to the previously studied MZM fusion in the non-interacting single particle picture, 
here we do not find any density fluctuations during the non-trivial fusion of MZMs (in one-dimensional geometry),
 using the time-evolving many-body wave-function. 
The formation of electron and hole  clearly appears in the electron and hole parts of the time-dependent
local-density of states. The local-density of states can be measured in tunneling-spectroscopy in quantum-dot experiments.

The study of non-trivial fusion is related to non-Abelian statistics and can be performed in quantum-dot setups. 
For successful fusion experiments, it is crucial to carefully consider the effects of interactions, disorder, and system size, 
because trivial low-energy modes could produce false positive results~\cite{Clarke, Danon}.
Our results indicate that even small inter-site Coulomb interactions can be destructive for the 
Poor Man's Majorana Zero Modes, 
causing all four MZMs to move away from zero energy for short chains systems.
Fortunately, such effect is absent for longer chains, showing that the topological protection increases with system size.

Our time-dependent results for the electron and hole components of the local density-of-states suggest 
that separately measuring the electron and hole portions of such local density-of-states 
could aid in distinguishing the fusion outcomes from trivial low-energy modes with unequal 
superposition of electron and hole states.  
However, the tunneling experiments generally measure the total local density-of-states, making it challenging to 
differentiate between fusion outcomes, low-energy modes, and topological MZMs.
Fortunately, 
recent advancements in quantum-dot systems provide a well-controlled setup, and the effective local gate control of 
individual quantum dots significantly reduces the effect of disorder or low-energy modes, 
making it feasible to measure MZMs more accurately than in nanowire systems~\cite{Mazur}. 
We believe our novel theoretical results for the non-trivial fusion of Majoranas in the one-dimensional chain 
and in the $Y$-shape geometries can be  realized
using the recently developed quantum-dot setups. Our prediction of the fusion outcome, 
based on the time-dependent local density-of-state method,
is accessible to the present-day experimental capabilities~\cite{Dvir}.
Braiding experiments using $Y$-shaped~\cite{pandey1} or $X$-shaped~\cite{pandey3} quantum dots 
should provide more conclusive and definitive evidence of MZMs and their non-Abelian properties, but such
efforts will be pursued in the future~\cite{Danon}.

\section {Acknowledgments}
The work of B.P., S.O., and E.D. was supported by the U.S. Department of Energy (DOE), Office of Science, Basic Energy Sciences (BES),
 Materials Sciences and Engineering Division.

    \section{APPENDIX A: Non-trivial fusion between a pair of quasi-MZMs and a pair of MZMs.}                                            
                                                                                                                                      
    \begin{figure}[h]                                                                                                                 
    \centering                                                                                                                        
     \rotatebox{0}{\includegraphics*[width=\linewidth]{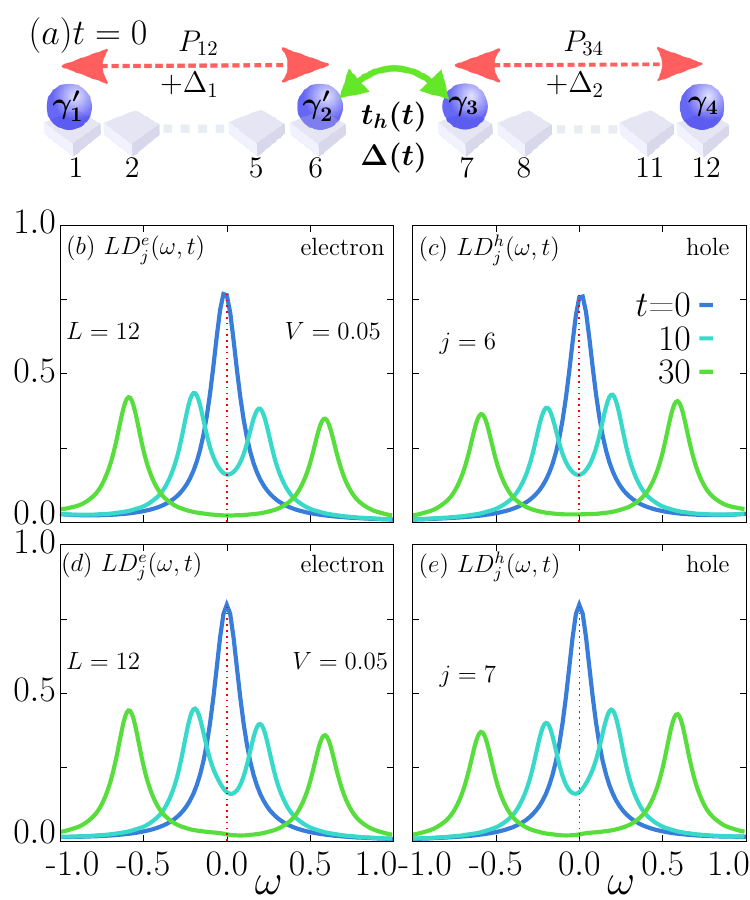}}
      \caption{(a) Schematic representation of a 12-sites 
quantum dots array with different left and right 
superconducting pairing strength $\Delta_1=0.7$ and $\Delta_2=1.0$. The left part has overlapping
      quasi-MZMs ($\gamma'_1$, $\gamma'_2$) as the system is away from the sweet-spot. The right part has non-overlapping MZMs 
      ($\gamma_3$, $\gamma_4$) and has $t_h=\Delta_2=1$.   
      At time $t=0$, there is no hopping and no pairing coupling between the Majoranas $\gamma'_2$ and $\gamma_3$. 
  (b) The electron component  $LD_j^e(\omega,t)$ and (c) the hole component $LD_j^h(\omega,t)$,
at site $j=6$ and for different times $t$, all at $V=0.05$. 
  (d) The electron component  $LD_j^e(\omega,t,j)$ and (e) the hole component $LD_j^h(\omega,t)$,
at site $j=7$ and for different times $t$, all at $V=0.05$. 
      }     
\label{fig9}
    \end{figure}                                                                                                                      
In Fig.~\ref{fig9}, we present the non-trivial fusion for 1D quantum dots with different pairing strengths $\Delta_1=0.7$
and $\Delta_2=1.0$. As the left part is away from the sweet-spot, the MZMs ($\gamma'_1$ and $\gamma'_2$) spatially overlap
with each other and move slightly away from zero energy [see  Fig.~\ref{fig9} (b,c)]. On the other hand, the
right MZMs pair ($\gamma_3$ and $\gamma_4$) are at sweet spot $\Delta_2=t_h=1$  and they do not overlap with each other.
For the pairs on the right, we find equal-height peaks at $\omega=0$ in the  electron and hole parts 
of the local density-of-states
at $t=0$ [Fig.~\ref{fig9} (d,e)]. This is expected for the MZMs as they are their own antiparticles, 
with equal contribution of electron and hole components~\cite{Herbrych}. For $t>0$, we vary the time-dependent hopping
and pairing terms between the left and right quantum dot segments, in order to fuse the quasi-MZM $\gamma_2'$ (at $j=6$) 
and MZM $\gamma_3$ (at site $j=7$) non-trivially, as described in the main text. 
Interestingly, in this case with increasing time, we find unequal-height electron and hole peaks close to 
$\omega=\pm 2t/\tau$. The unequal peak height comes from the fact that for this case the time-evolving wavefunction 
$|\Psi(t)\rangle$ has an unequal overlap with the low-energy states $|\Psi_1\rangle$ and $|\Psi_4\rangle$, which is quite different from the non-trivial fusion, where the left and right pairs of MZMs
were initialized at the sweet spot $\Delta_1=\Delta_2=1$. This shows that our time-dependent results for the electron and hole components of the local density could be used to distinguish 
the  fusion outcomes from trivial low-energy modes (which appears
after hybridization of two MZMs in the same region).  However, we find that the total local density-of-states 
at $\omega=\pm 2t/\tau$  shows very similar behavior,  as compared to the previous case of non-trivial fusion.

\section{APPENDIX B:  Analytic calculation of local density-of-states for a four-site Kitaev Hamiltonian}                                   
     \begin{figure}[h]                                                                                                                 
     \centering                                                                                                                        
     \rotatebox{0}{\includegraphics*[width=\linewidth]{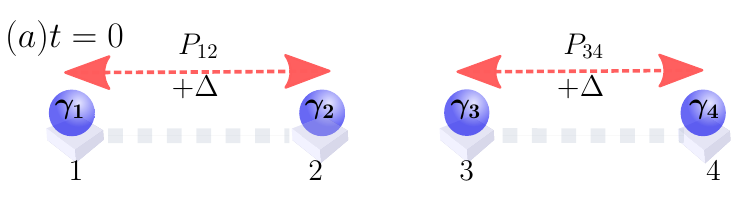}}
     \caption{Schematic representation of two pairs of MZMs using a four-sites quantum-dots array 
with same left and right             
     superconducting phases $\phi_1=\phi_2=0$.}                                                                                        
     \label{fig5}                                                                                                                     
     \end{figure}                                                                                                                      
                                                                                                                                       
  In this Appendix, we calculate analytically results for the four-site quantum-dots at the                                       
    sweet spot ($t_h=\Delta=1$) to find the exact peak positions                                                                      
    during the fusion process. The four-site Kitaev model without any coupling between the left and 
    right quantum dot arrays can be written as:
     \begin{eqnarray}
     H^L =  -t_hc^{\ {\dagger}}_{1}c_{2} +  \Delta  c_1 c_{2} + H.c. ,\\
     H^{R} = -t_hc^{\ {\dagger}}_{3}c_{4} + \Delta  c_3 c_{4} + H.c. .
     \end{eqnarray}
   
    In the many-body basis ($2^4=16$ states), this four-site Hamiltonian is four-fold degenerate
    with two states having total even parity, and the other two having odd parity.
    The four degenerate eigenvalues at $t_h=\Delta$ take the form $E_n=-2\Delta$.
    We have a total of four Majorana zero modes localized at each site, as shown in  Fig.~\ref{fig5}.
    The ground state-wave function in the presence of a very small interaction $V=0.001$ (such that we have well-defined total parity) is
    \begin{equation}
    |\Psi_0\rangle = 0.5|0101\rangle +0.5 |0110\rangle + 0.5|1001\rangle + 0.5 |1010\rangle.
    \end{equation}
    Note that the total fermion number parity of the system $P=exp({i\pi\sum_jn_j})=+1$ (even),
    but the individual parity of left and right quantum-dot segment is odd (as state kets take the form $|01\rangle$ $\otimes$ $|10\rangle$).
   
     Next, we introduce the hopping and pairing coupling between the left and right quantum dots as a parameter:
    \begin{eqnarray}
          H^{C} =  -t^h_{23} c^{\ {\dagger}}_{2}c_{3} +  \Delta_{23}  c_2 c_{3} + H.c. .
    \end{eqnarray}
   
     For finite values of $t^h_{23}= \Delta_{23}$ the four-fold degeneracy of the system splits into two pairs $\left[(E_1, E_2) \text{and} (E_3, E_4)\right]$ of two-fold  degenerate states with combination of even and odd fermion parity.
     The four lowest eigenvalues are
 \begin{eqnarray}                                                                                                                  
    E_1=-2\Delta -\Delta_{23},\\                                                                                                      
    E_2=-2\Delta -\Delta_{23},\\                                                                                                      
    E_3=-2\Delta +\Delta_{23},\\                                                                                                      
    E_4=-2\Delta +\Delta_{23}.                                                                                                        
    \end{eqnarray}                                                                                                                    
    We found that the ground state wave-function for finite values of $\Delta_{23}=0.2$ (in the units of $t_h$ unless otherwise stated)
    (and very small values of $V=0.001$)
     can be written as                                                                                                                 
     \begin{eqnarray}                                                                                                                  
     |\Psi_0\rangle =0.353 |0000\rangle+ 0.353 |0011\rangle+ 0.353|0101\rangle  \nonumber\\ +0.353 |0110\rangle 
       + 0.353|1001\rangle + 0.353 |1010\rangle  \nonumber\\ + 0.353 |1100\rangle+  0.353 |1111\rangle .
     \end{eqnarray} 
 Interestingly, now for the ground state $|\Psi_0\rangle$, the left and right part of the quantum-dot segments have both even and odd parity
     kets, which allow for the formation of both electron and hole components in equal magnitude.                                      
     The electron part of the local density-of-states at site $j$ can be calculated as:                                                
     \begin{equation}                                                                                                                  
             LD_j^{e}(\omega,j)= -\frac{1}{\pi}Im\left(\sum_m\frac{\langle\Psi_1|c^{\dagger}_j|\Psi_m \rangle  \langle\Psi_m|c_j|\Psi_1 \rangle }{\omega+E_m-E_1+i \eta}\right).
     \end{equation}                                                                                                                    
     We find that the peak in the electron or hole portions of the local density-of-states arise from the                              
     transition between the states ($m=3$ and $m=1$) with opposite fermion parity.                                                     
     The peak position for $\Delta_{23}=0.2$  in the electron part of the local density-of-states appears                              
     at $\omega= E_1-E_3=-2 \Delta_{23}=-0.4$.                                                                                         
    Similarly for the hole part of the local density-of-states, the peak appears at $\omega= E_3-E_1=2 \Delta_{23}=+0.4$.             
    For the static case in the non-trivial fusion process,                                                                            
    these results show that we have the formation of an electron at $\omega=-2\Delta_{23}$ and a hole at  $\omega=+2\Delta_{23}$.     
    The peak height in the electron and hole parts of the local density-of-states are almost the same,
    indicating the formation of equal magnitude of electron and hole 
    in the non-trivial fusion process~\cite{souto}.

    \section{APPENDIX C: The overlap of time-evolving wavefunctions with other low-energy states}
     \begin{figure}[h]                                                                                                                 
     \centering                                                                                                                        
     \rotatebox{0}{\includegraphics*[width=\linewidth]{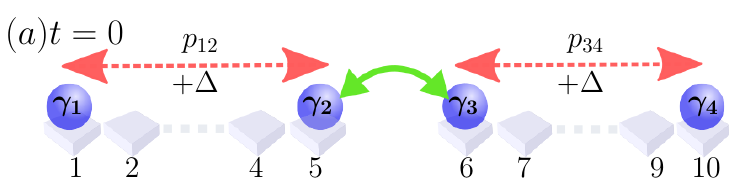}}
     \rotatebox{0}{\includegraphics*[width=\linewidth]{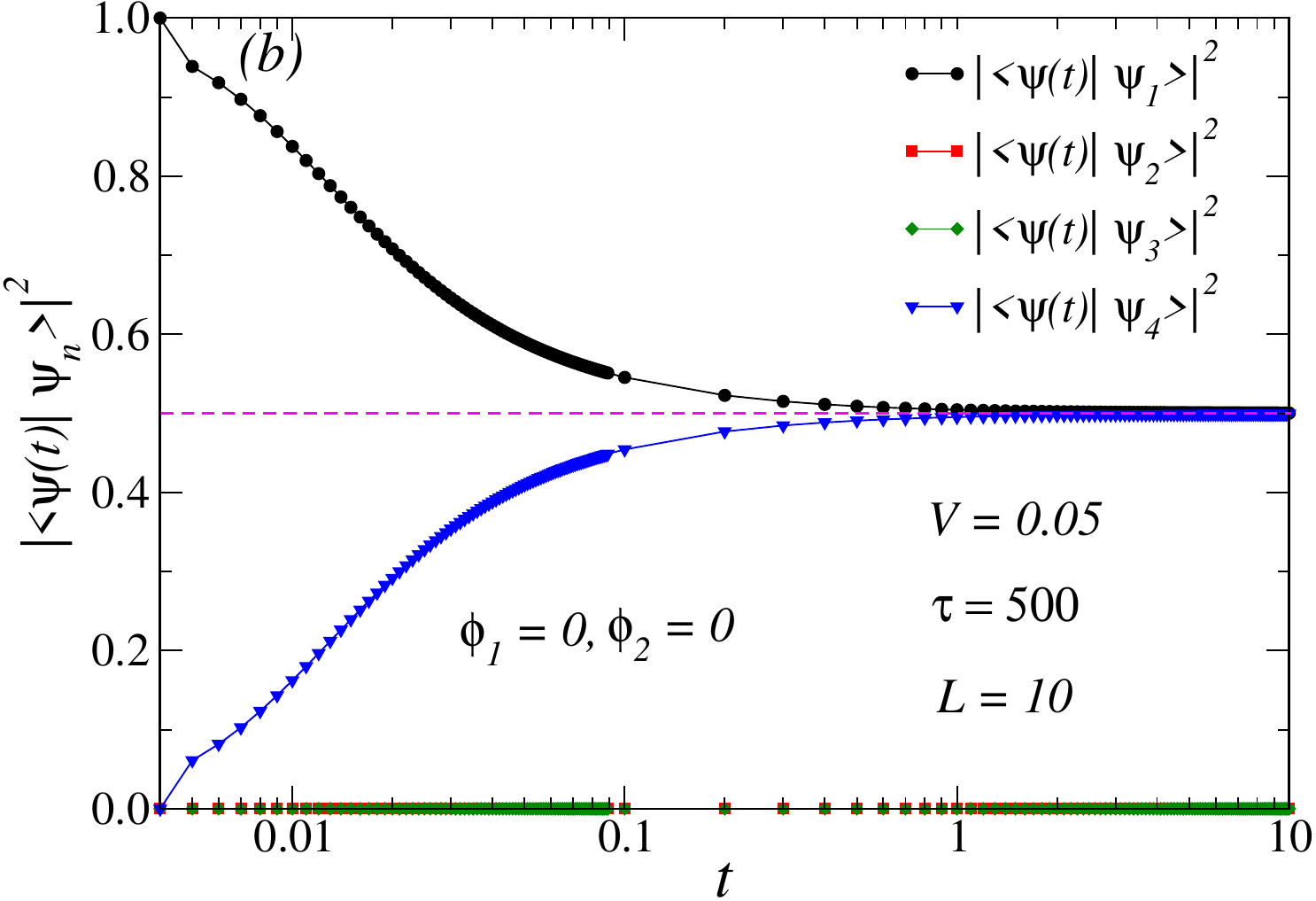}}
    \caption{(a) Schematic representation of two pairs of MZMs using a 10-sites quantum-dots array with same left and right           
    superconducting phases $\phi_1=\phi_2=0$. At time $t=0$, there is no hopping and no pairing coupling between the Majoranas 
       $\gamma_2$ and $\gamma_3$. (b) The overlap of time-evolving wavefunction $\Psi(t)$ with
     the four lowest states of the instantaneous Hamiltonian at time $t$ (in log scale).                                              
    The numerical calculations were performed using the full-diagonalization of a $L=10$ sites cluster                                
    and $V=0.05$, $\tau=500$.}                                                                                                        
    \label{fig6}                                                                                                                     
    \end{figure}                                                                                                                      
     \subsection{1. $\phi_1=\phi_2=0$.}
                                                                                                                                      
    In  Fig.~\ref{fig6}, we calculate the overlap of the time-evolving states $|\Psi(t)\rangle$ with                                
    the four lower states of the instantaneous                                                                                        
    Hamiltonian $H(t)$ in the processes of non-trivial fusion of MZMs. We start the time evolution using the ground state             
    $|\Psi(0)\rangle$ with total parity $P=+1$ of the system shown in  Fig.~\ref{fig6}(a), where both the                           
    left and right parts of the quantum-dots                                                                                          
    arrays have the same superconducting phases $\phi_1=\phi_2=0$ and $t_h=|\Delta|=1$. At time $t=0$, there is no hopping and no pairing coupling between
    the left and right quantum-dot segments (Fig.~\ref{fig6}(a)).     
    As described in the main text, we perform the time evolution by changing the hopping according to the formula
    $t_h(t)=t_h(T)\frac{n\delta t}{\tau}$ using a system size $L=10$ sites and $\tau=500$. As shown in Fig.~\ref{fig6}(b),
    the  time evolving states $|\Psi(t)\rangle$ becomes an equal superposition of two states $\Psi_1$ and   $\Psi_4$ very shortly with increase in time.
    Note that the individual total parity of the  $\Psi_1$ and $\Psi_4$ states are the same as in the case of
    the time-evolving states  $|\Psi(t)\rangle$.
     The equal  superposition of these two states $\Psi_1$ and   $\Psi_4$, results in equal amount of spectral weight at $\omega=\pm 2t/\tau$
     in both electron and hole part of the local-density of states.

     \subsection{2.  $\phi_1=\pi$ and $\phi_2=0$.}

     \begin{figure}[h]
     \centering
     \rotatebox{0}{\includegraphics*[width=\linewidth]{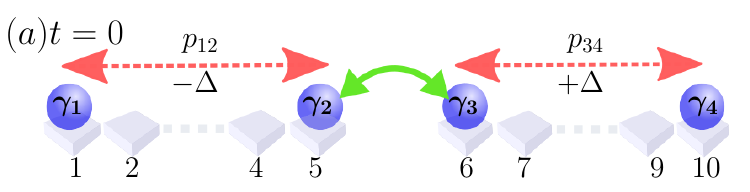}}
     \rotatebox{0}{\includegraphics*[width=\linewidth]{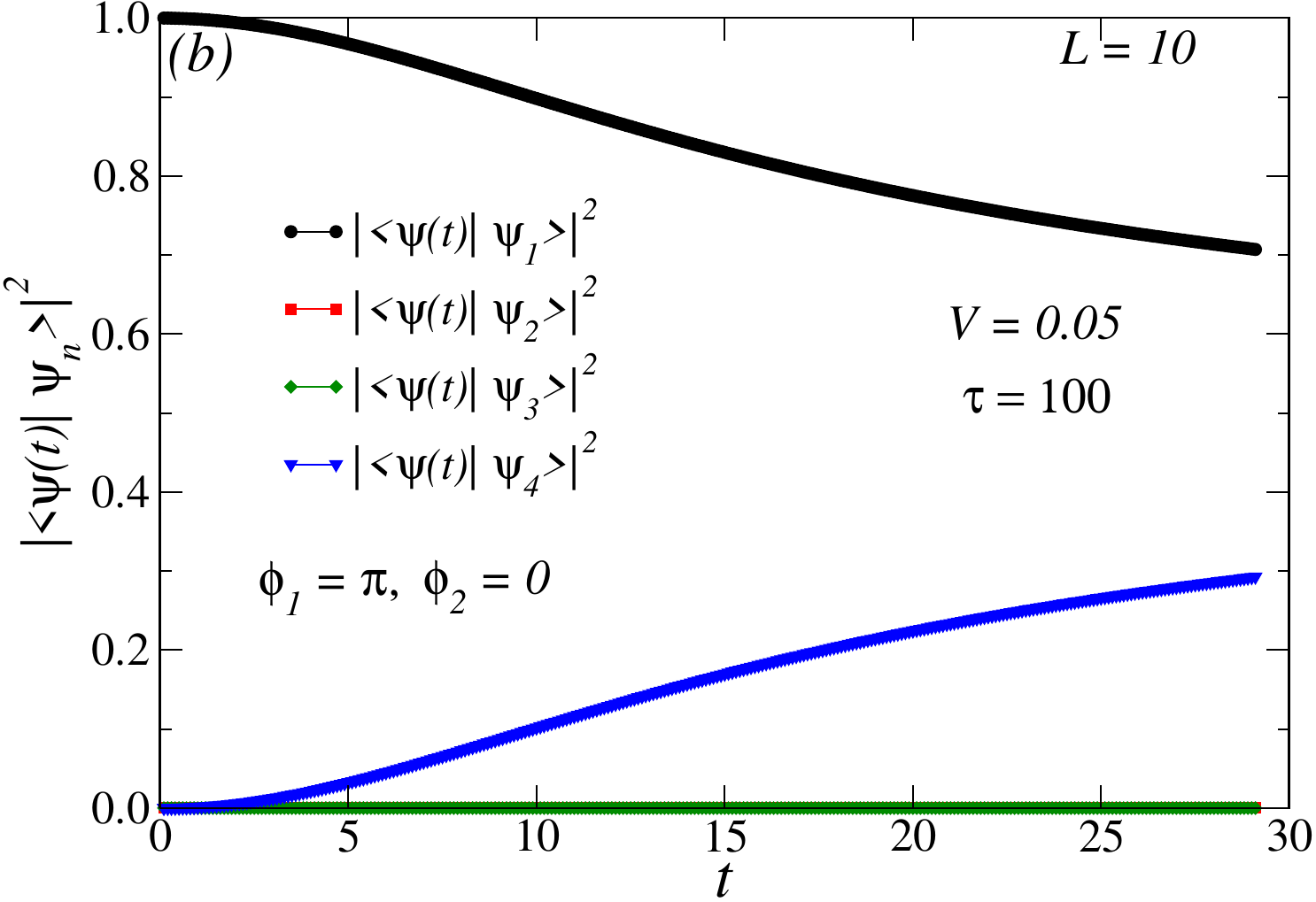}}
     \caption{(a) Schematic representation of two pairs of MZMs using a 10-sites quantum dots array with different left and right
     superconducting phases $\phi_1=\pi$ and $\phi_2=0$. At time $t=0$, there is no hopping and no pairing coupling between the Majoranas $\gamma_2$ and $\gamma_3$.
     (b) The overlap of the time-evolving wavefunction $\Psi(t)$ with the four lowest states of the instantaneous Hamiltonian at time $t$.
     The numerical calculations were performed using full-diagonalization of $L=10$ sites and $V=0.05$, $\tau=100$.}
     \label{fig7}
    \end{figure}
    In  Fig.~\ref{fig7}, we consider a $\pi$-junction quantum-dot array, where both the left and right parts
    of the system have different superconducting phases $\phi_1=\pi$ and $\phi_2=0$. The parameters are $t_h=|\Delta|=1$.
     We start the time evolution using the ground state $|\Psi(0)\rangle$ with total parity $P=+1$ of the system shown in  Fig.~\ref{fig7}(a).
     At time $t=0$, there is no hopping and no pairing coupling between the left and right quantum-dot segments (Fig.~\ref{fig7}(a)).
    Next, we calculate the overlap of the time-evolving sites, using $\tau=100$.
    As shown in Fig.~\ref{fig7}(b), the  time evolving states $|\Psi(t)\rangle=u|\Psi_1\rangle+ v |\Psi_4\rangle$
     becomes a superposition of two low-energy degenerate ground states (with the same total parity).
    The amplitude $u$ and $v$ depends on  the tunneling of the  MZM $\gamma_3$ from site $j=6$ to
    site $j=4$ and changes far more slowly as compared to the case when  $\phi_1=\phi_2=0$. Interestingly, the system
    remains four-fold degenerate even with the change with time in $t_h(t)=\Delta(t)$ at the bond between
    site $j=5$ and $6$.
    This result shows that in the case of the tunneling of MZM through a localized Majorana at site $j=5$,
    it still leads to a superposition of two low-energy states. Note that the amplitudes $u$ and $v$ remain the same for different values of large $\tau$.

    \section{APPENDIX D: Local density-of-states for different values of the quench rate $1/\tau$.}                                            
                                                                                                                                      
    \begin{figure}[h]                                                                                                                 
    \centering                                                                                                                        
     \rotatebox{0}{\includegraphics*[width=\linewidth]{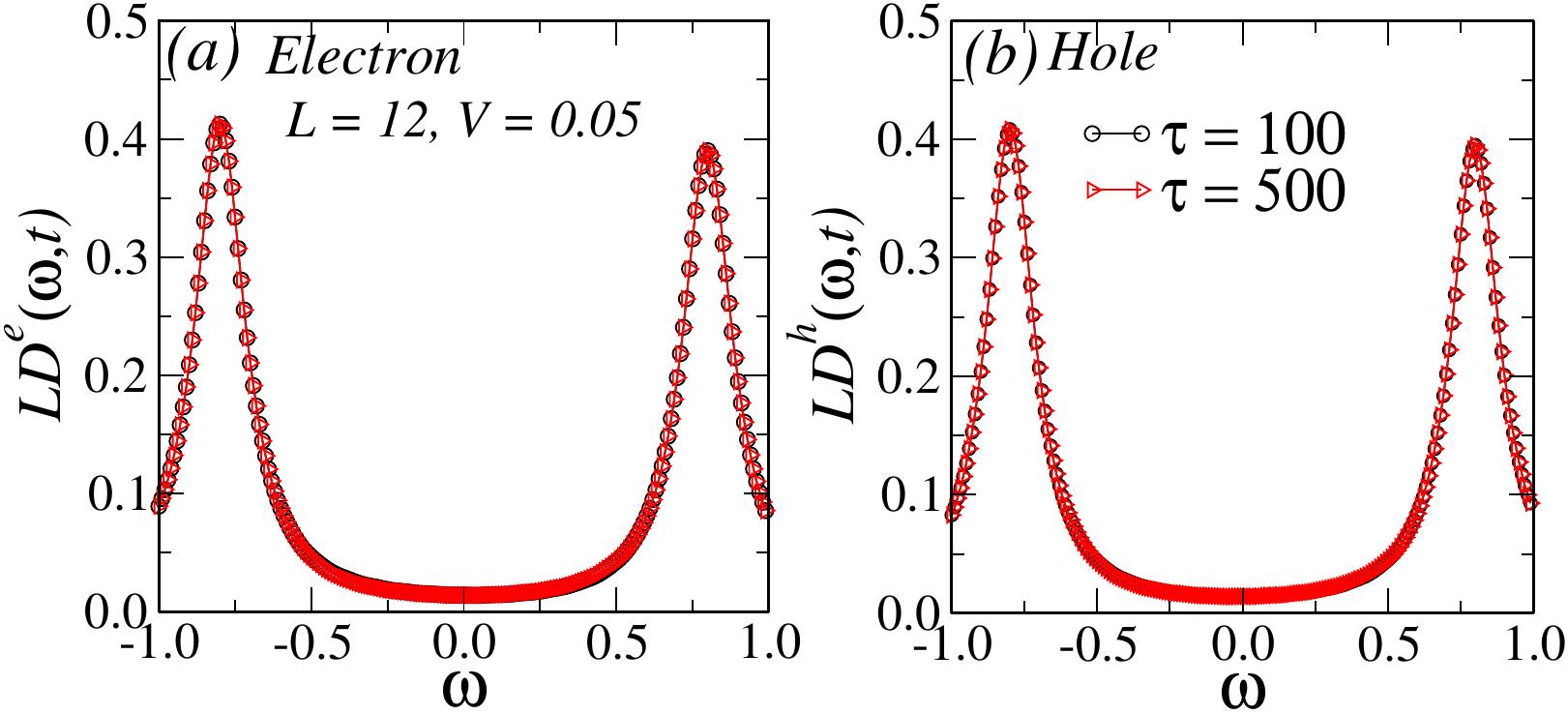}}
    \caption{Comparison of the local density-of-states at $t/\tau=0.4$ and site $j=6$, and for different values of $\tau=100$ and $500$:
    (a) electron part of the local density-of-states $LD^e_j(\omega,t)$, and (b) hole part of the local density-of-state  $LD^h_j(\omega,t)$.
    These numerical calculations were performed using $L=12$ sites and $V=0.05$.}                                                     
    \label{fig8}                                                                                                                     
    \end{figure}                                                                                                                      
    In the main text of the paper, for the time evolution of the one-dimensional quantum-dots arrays, we used $\tau=100$,             
     which is already a quite small quench rate $1/\tau$.                                                                             
    To confirm that this value characterizes an adiabatic process, in Fig.~\ref{fig8} we have compared the electron and hole parts of the local density-of-states
    at  $t/\tau=0.4$ for two different values of $\tau$ i.e.                                                                          
    $\tau=100$ and $500$. As shown in  Fig.~\ref{fig8}(a), the electron                                                             
    part of the local density-of-states at $\omega=\pm 2t/\tau=\pm 0.8$ remain the same for two different                             
     values of  $\tau=100$ and $500$.                                                                                                  
     Similarly, we also find the hole part of the  local density-of-states also take the                                               
     same values for two different values of  $\tau=100$ and $500$.                                                                    
     These results shows that the quench rate $1/\tau$ considered in the main text ($\tau=100$) is already a quite small               
     number, representative of adiabatic movement,                                                                                     
     and the dynamical results are unchanged even for larger values of $\tau$. 



\end{document}